\documentclass[manuscript]{aastex}

\usepackage{graphicx}
\usepackage{rotating}
\usepackage{dcolumn}

\newcommand{\rsun}{R_{\Sun}}
\newcommand{\mydeg}{^{\circ}}

\shorttitle{2008 Dec 12 CME Deflection}
\shortauthors{Kay et al.}

\begin{document}

\title{Constraining the Mass and the Non-Radial Drag Coefficient of a Solar Coronal Mass Ejection}

\author{C. Kay}
\affil{Astronomy Department, Boston University, Boston, MA 02215}
\email{ckay@bu.edu}

\author{L. F. G. dos Santos}
\affil{Physics Department, Universidade de Bras\'{i}lia, Bras\'{i}lia, DF, 70910-900, Brazil}

\and 

\author{M. Opher}
\affil{Astronomy Department, Boston University, Boston, MA 02215}

\begin{abstract}
Decades of observations show that CMEs can deflect from a purely radial trajectory yet no consensus exists as to the cause of these deflections.  Many of theories attribute the CME deflection to magnetic forces.  We developed ForeCAT \citep{Kay13, Kay14}, a model for CME deflections based solely on magnetic forces, neglecting any reconnection effects.  Here we compare ForeCAT predictions to the observed deflection of the 2008 December 12 CME and find that ForeCAT can accurately reproduce the observations.  Multiple observations show that this CME deflected nearly 30$\mydeg$ in latitude \citep{Byr10, Gui11} and 4.4$\mydeg$ in longitude \citep{Gui11}.  From the observations, we are able to constrain all of the ForeCAT input parameters (initial position, radial propagation speed, and expansion) except the CME mass and the drag coefficient that affects the CME motion.  By minimizing the reduced chi-squared, $\chi^2_{\nu}$, between the ForeCAT results and the observations we determine an acceptable mass range between 4.5x10$^{14}$ and 1x10$^{15}$ g and the drag coefficient less than 1.4 with a best fit at 7.5x10$^{14}$ g and 0 for the mass and drag coefficient. ForeCAT is sensitive to the magnetic background and we are also able to constrain the rate at which the quiet sun magnetic field falls to be similar or to or fall slightly slower than the Potential Field Source Surface model.
\end{abstract}

\keywords{Sun: coronal mass ejections (CMEs)}

\section{Introduction}
Predicting the deflections of coronal mass ejections (CMEs) is key for forecasting space weather effects.  CME deflections can occur in the corona or interplanetary space \citep{Kil09, Isa13, Rod11, Wan02, Wan04, Wan06, Wan14}, or as the result of CME-CME interactions \citep{Lug12}.  Coronal deflections have long been associated with the presence of coronal holes \citep{Cre06, Gop09, Moh12}, however, more generally, CMEs may be deflected by gradients in the background solar magnetic field \citep{Gui11, She11} which tends to deflect CMEs towards the Heliospheric Current Sheet (HCS) on global scales.
 
We develop a model, Forecasting a CME's Altered Trajectory (ForeCAT) \citep{Kay13, Kay14}, for coronal CME deflections determined by the background solar magnetic field which can, in general, produce CME deflections of comparable magnitude to observations.  \citet{Kay14} use an improved version of ForeCAT to show that magnetic deflections follow observed solar cycle trends: primarily latitudinal deflections occur near solar minimum but more variety in direction as well as larger magnitudes occur as the HCS becomes more inclined and the background magnetic field strength increases.  In \citet{Kay14}, we find that a CME typically obtains a constant deflection velocity in the lower corona and continues deflecting through interplanetary space as the drag and magnetic forces cannot significantly reduce the CME's motion when the drag coefficient is of order unity.  Deflections in interplanetary space proportional to the inverse of the radial distance distance are the signature of a constant radial velocity and a constant deflection velocity and do not require additional interplanetary deflection forces \citep{Kay14}. 

In this letter, we present the first comparison between ForeCAT model results and an observed CME.  Observations determine most of the initial inputs for ForeCAT (initial location and shape, and propagation and expansion models), leaving only the CME mass and solar wind drag coefficient as free parameters.  We show that not only can ForeCAT reproduce the observed deflection, but it also constrains the unknown mass and drag coefficient as well as the background magnetic field.

\section{ForeCAT}
ForeCAT simulates CME deflections from the forces determined from a static background magnetic field \citep{Kay14}.  The flux rope of a CME is represented as a toroidal structure.  To initialize a CME, ForeCAT requires the CME position (latitude, longitude, and tilt with respect to the solar equatorial plane), shape (height, width, and cross-sectional width), and mass.     

The most important free parameters relate to the propagation and expansion of the CME.  ForeCAT uses a three-phase radial propagation model similar to that presented in \citet{Zha06}.  The CME begins in a slow rise phase, followed by rapid acceleration, and finally a constant propagation phase.  \citet{Kay14} assume the simplest form of expansion, self-similar expansion, however, ForeCAT can utilize other expansion models based on constraints from observations.

The forces responsible for the deflection, both magnetic pressure gradients and magnetic tension, are calculated at grid points across the face of the CME torus.  The net force determines the deflection motion.  ForeCAT includes a non-radial drag force which restricts the deflection.  We use the standard hydrodynamic form of drag \citep{Car96, Car04} with a drag coefficient proportional to $\tanh \beta$ \citep{For06} and assume purely radial solar wind.  The volumetric drag is calculated as

\begin{equation} \label{drageq}
F_D = -\frac{2 C_d \tanh \beta \rho_{SW}}{\pi b} \vec{v}_{CME,nr} | \vec{v}_{CME,nr} | 
\end{equation}

where $C_d$ is the drag coefficient, $b$ is the CME's cross-sectional width, $\rho_{SW}$ is the solar wind density, and $\vec{v}_{CME,nr}$ is the non-radial CME velocity.  As in \citet{Kay14}, we ignore the factor of less than unity which describes the projection of the cross-sectional area onto the propagation direction. We use an empirical form for $\beta$ determined from Figure 1.22 of \citet{AscB}: $\beta (R) = 2.515 (R -1)^{1.382}$, with $R$ being the heliocentric radial distance.

ForeCAT uses simplified models for the background solar wind.  The solar magnetic field is represented using a Potential Field Source Surface (PFSS) model \citep{Alt69, Sch69, Alt77} with the Parker interplanetary magnetic field included above the source surface.  ForeCAT describes the solar wind density using a modified version of the \citet{Guh06} model in which the density is the sum of a current sheet model and a coronal hole model, with the relative weight of the contributions determined by the distance from the HCS.  We vary the background to explore the effect of different magnetic field profiles, which we discuss later.

\section{Observations}\label{Obs}
While the 2008 Dec 12 has been extensively covered in the literature, we focus on the results of \citet{Byr10} and \citet{Gui11}, hereafter B10 and G11, as they offer the most complete coverage.  Figure \ref{fig:bestfit} shows the latitude, Stonyhurst longitude (G11 only), and radial velocity of the CME nose out to 50 $\rsun$ for both the B10 data (blue squares) and the G11 data (red circles).  Both authors utilize data from the Sun-Earth Connection Coronal and Heliospheric Investigation (SECCHI) on board the twin Solar Terrestrial Relations Observatory (STEREO) satellites.  At the time of the eruption the satellites were separated by 86.7$\mydeg$.  

Fig. \ref{fig:bestfit} includes error bars estimated from the scatter in the G11 and B10 observations.  For the latitudinal data, the error bars represent the absolute difference between the two sets of measurements at each height.  We find average deviations of 0.5$\mydeg$ below 3 $\rsun$, 2$\mydeg$ between 3 and 10 $\rsun$, and 4$\mydeg$ beyond 10 $\rsun$.  For the longitudinal data, the uniform error bars of 4$\mydeg$ are determined from the scatter within the G11 data.  

\section{Determination of ForeCAT Inputs}
ForeCAT CME's are initiated at a height of 0.05 $\rsun$ above photospheric polarity inversion lines (PILs).  Using a finite number of harmonic coefficients causes a ringing effect in the PFSS model, and for coefficients up to order 90 this effect disappears by 1.05 $\rsun$.  Using an Helioseismic and Magnetogram Imager (HMI) synoptic magnetogram for Carrington Rotation (CR) 2077 we identify a PIL in the quiet sun (not associated with an active region) centered at 52$\mydeg$ latitude and 81.7$\mydeg$ longitude with a tilt of -13.8$\mydeg$, near the initial location of the observations.   

B10 find that the empirical relationship $w(R) = 13R^{0.22}$ describes the CME's angular width, $w$ in degrees, as a function of radial distance, $R$ in solar radii, which we use to determine the CME width, $c$.  Both the CME height and cross-sectional width (the parameters $a$ and $b$ in \citet{Kay14}) are set at constant fractions of the CME width ($a=Ac$ and $b=Bc$).  The fraction $A$ is estimated from the radial distance between the flanks and the nose of B10 and is set at one throughout the duration of the CME's propagation.  The cross-sectional radius of the CME is harder to constrain.  We set the cross-sectional radius fraction $B=\frac{1}{4}$ and explore the effects of these chosen values in section \ref{shapeparams}.

The bottom panel of Fig. \ref{fig:bestfit} shows the observed radial velocity determined from the change in position of the CME nose, which combines the effects of expansion and propagation.  Since we know the expansion, matching ForeCAT's nose velocity to these observations allows us to constrain the radial propagation model.  ForeCAT initiates the CME with a radial propagation speed of 75 km s$^{-1}$.  Once the CME reaches 1.75 $\rsun$ it begins constantly accelerating until it reaches a final propagation speed of 480 km s$^{-1}$ at 18 $\rsun$.  Beyond this distance, the CME moves with constant speed.

\section{ForeCAT Results}
\begin{figure}
\includegraphics[scale=0.8]{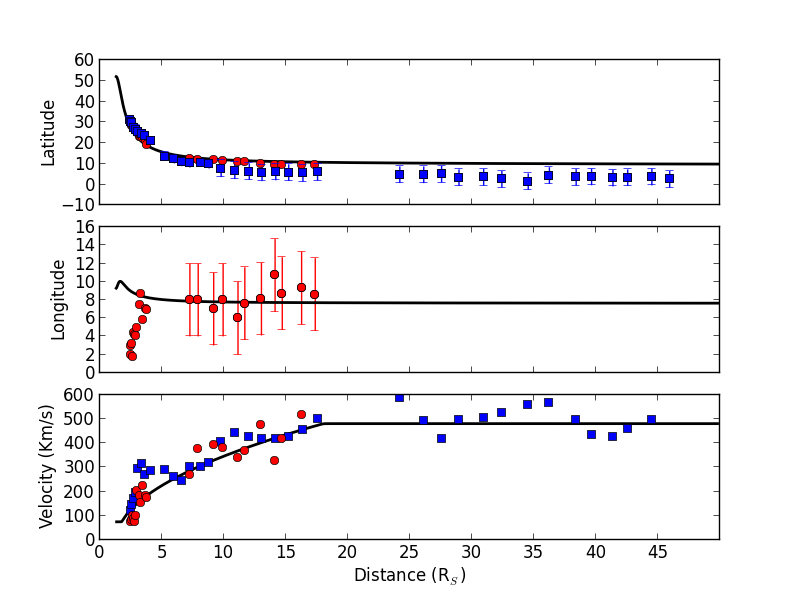}
\caption{Comparison of ForeCAT results with observations of latitude (top panel) and longitude (middle panel) versus distance for the 2008 December 12 CME.  The bottom panel shows the radial velocity of the CME nose, which combines propagation and expansion, versus radial distance which is used to constrain the propagation model.  The black line represents the best fit from ForeCAT, the blue squares represent the results of \citet{Byr10}, and the red circles the results of \citet{Gui11}.}\label{fig:bestfit}
\end{figure}

We constrain all of the ForeCAT input parameters from the observations except for the CME mass and drag coefficient and vary these parameters to obtain a best fit with the observed deflection.  Typical CME masses range between 10$^{14}$ and 10$^{16}$ g \citep{Gop09b} but observations suggest that the 2008 Dec 12 CME had a mass less that 2x10$^{15}$ g below 10 $\rsun$ \citep{Car12,DeF13}.  The drag coefficient, $C_d$, is typically set near unity, but \citet{Car04} determine coefficients as high as 300 for individual simulations.  Fig. \ref{fig:bestfit} shows ForeCAT results for a CME with a mass of 7.5x10$^{14}$ g and drag coefficient set to 0, which corresponds to the reduced chi-squared best fit to the G11 measurements, which we discuss in section \ref{Chi2}.  

Fig. \ref{fig:bestfit} shows that ForeCAT reproduces the latitudinal motion of the CME.  We also find decent agreement for the longitudinal motion beyond 5 $\rsun$.  The longitudinal motion below 5 $\rsun$ does not match the G11 data, however, measurements of longitudinal deflections in the low corona are inherently highly uncertain as line-of-sight coronagraph obsevations integrate in the longitudinal direction, so we do not include these points when determining a best fit.

We see a change in the direction of the longitudinal deflection below 2 $\rsun$.  As shown in \citet{Kay14}, the deflection is a combination of both local and global magnetic gradients.  The local gradients initially cause the CME to move very briefly westward.  This motion is small due to the weak local gradients of the solar minimum quiet sun.  As the CME propagates out radially the local gradients are overcome by global gradients determined by the relative location of the coronal holes and the HCS.  These global gradients are strong enough to slow down and change the CME's deflection to a eastward direction. 

\begin{figure}
\includegraphics[scale=0.2]{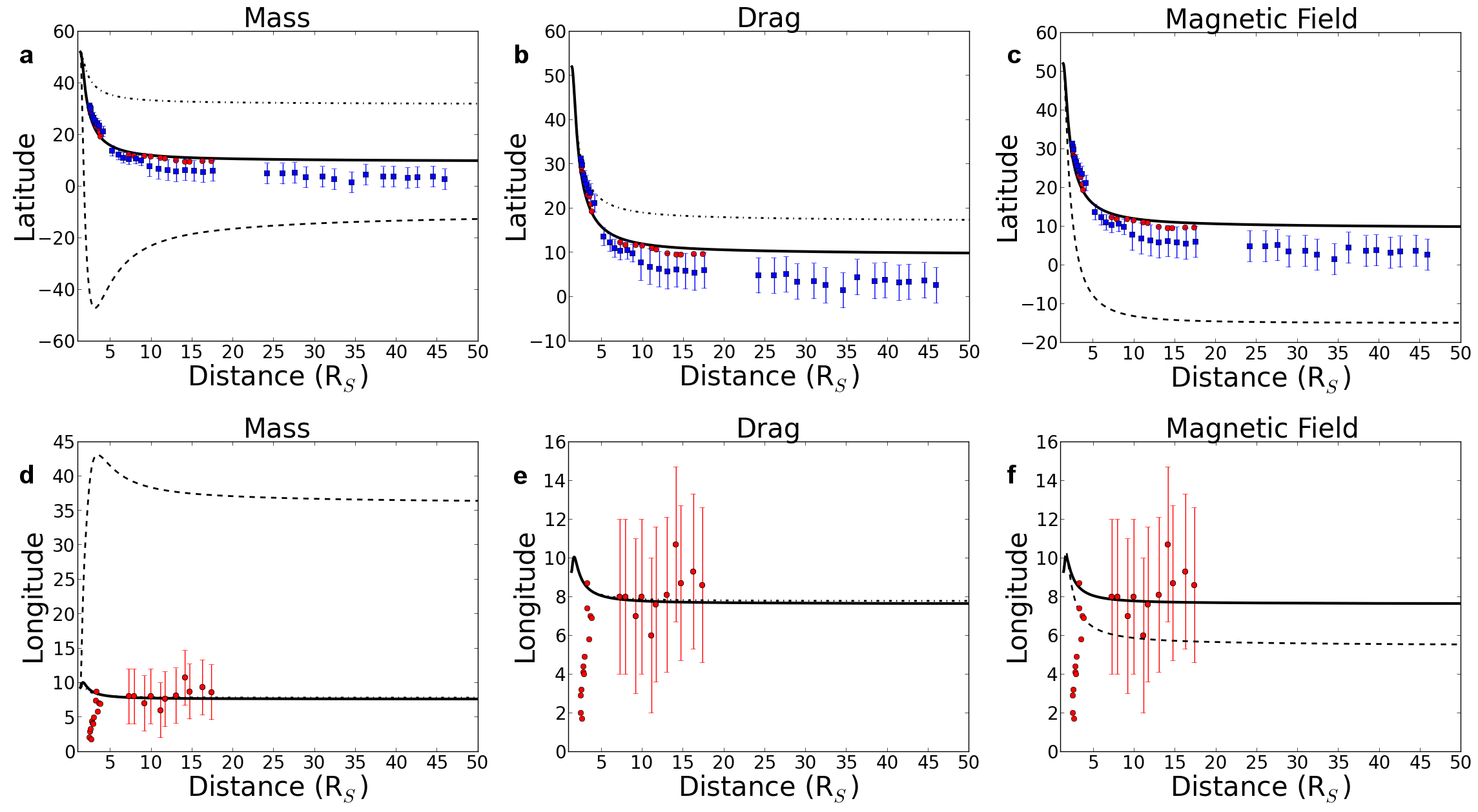}
\caption{Sensitivity of the latitudinal ((a)-(c)) and longitudinal ((d)-(f)) deflection to variations in the CME mass ((a) and (d)), drag coefficient ((b) and (e)), and background magnetic field model ((c) and (f)). The solid line represents the ForeCAT best fit and blue and red points the observations, as in Fig. \ref{fig:bestfit}.  In Panels (a) and (d) the dashed line corresponds to a mass of 10$^{14}$ g and the dot-dashed line corresponds to a CME mass of 2x10$^{15}$ g. In Panels(b) and (e) the dot-dashed line corresponds to a drag coefficient of 2.  In Panels \ref{fig:Lat}(c) and (f)the dashed line corresponds to the scaled PFSS magnetic field model.}\label{fig:Lat}
\end{figure}

\subsection{Variation with CME Mass}
Figure \ref{fig:Lat}(a) and (d) show the effect of varying the CME mass on the latitudinal and longitudinal deflections in the same format as Fig. \ref{fig:bestfit}.  The dashed lines represent a low mass CME case of 10$^{14}$ g and the dot-dashed lines represent a higher mass case of 2x$10^{15}$ g.  All three masses have similar latitudinal profiles with less massive CMEs deflecting more as the decrease in density allows for greater nonradial acceleration of the CME, as seen in \citet{Kay14}. The initial gradients deflect the low mass CME beyond the HCS but near 3 $\rsun$ the latitudinal deflection reverses direction, bringing it back toward the HCS.

Fig. \ref{fig:Lat}(d) shows very little difference in the longitudinal behavior of the high mass and best fit CMEs with the high mass CME deflecting slightly less both initially westward and back eastward.  The low mass case again shows a significant increase in the total deflection but retains the same westward then eastward motion.

\subsection{Variation with Drag Coefficient}
Figures \ref{fig:Lat}(b) and (e) show the effect of varying the drag coefficient ($C_d$ in Eq. \ref{drageq}).  The drag coefficient is very small in the low corona when the plasma beta is high so the effects of scaling the drag coefficients are not noticeable for the first few solar radii.  Beyond a few solar radii, we see that the increased drag causes a decrease in both the latitudinal and longitudinal deflection.  The effect is more noticeable in the latitudinal direction as the CME deflects more in that direction.  

\subsection{Magnetic Background}
Observations of Type II radio bursts suggest that the PFSS magnetic field may fall too rapidly with distance in the corona above active regions \citep{Man03, Eva08}.  \citet{Kay14} show that ForeCAT is sensitive to the rate at which the background magnetic field decreases with distance.  

The 2008 December 12 CME occurs in the quiet sun, rather than above an active region, but we still explore the sensitivity of the deflection to the magnetic background.  Fig. \ref{fig:bestfit} shows the unscaled PFSS model results in a good fit with observations, however, a different magnetic background could potentially result in a better fit with a different parameters.  Figure \ref{fig:Lat}(c) and (f) compares the unscaled PFSS best fit (solid line) with the results for the same drag coefficient and CME mass, but using a magnetic background corresponding the the PFSS model increased by a factor of $R$ below 2.5$\rsun$.

Both the latitude and longitude show that decreasing the rate at which the magnetic field decays with distance results in stronger deflections.  For the CME mass and drag coefficient of the unscaled best fit, the deflections exceed the observed deflections.  In section \ref{Chi2}, we explore whether a different set of CME mass and drag coefficient produces a good fit.

\subsection{Determination of the Best Fit}\label{Chi2}
We determine the mass and drag coefficient of the best fit by computing the reduced chi-squared, $\chi^2_{\nu}$, which measures the variation between the ForeCAT model and the observations.  We calculate $\chi^2_{\nu}$ as
\begin{equation}
\chi^2_{\nu} = \frac{1}{N-\nu-1} \Sigma \frac{(y_{obs} - y_{FC})^2}{\sigma^2_{obs}}
\end{equation}
where $N$ is the number of data, $\nu$ is the degrees of freedom, $y_{obs}$ are the observed position, $y_{FC}$ are the ForeCAT position, and $\sigma_{obs}$ are the uncertainty as defined in section \ref{Obs}.  Computing $\chi^2_{\nu}$ requires comparing $y_{obs}$ and $y_{FC}$ at the same radial distance so we linearly interpolate between the ForeCAT results to determine the ForeCAT values at the distances of the observations.

\begin{figure}
\includegraphics[scale=0.35]{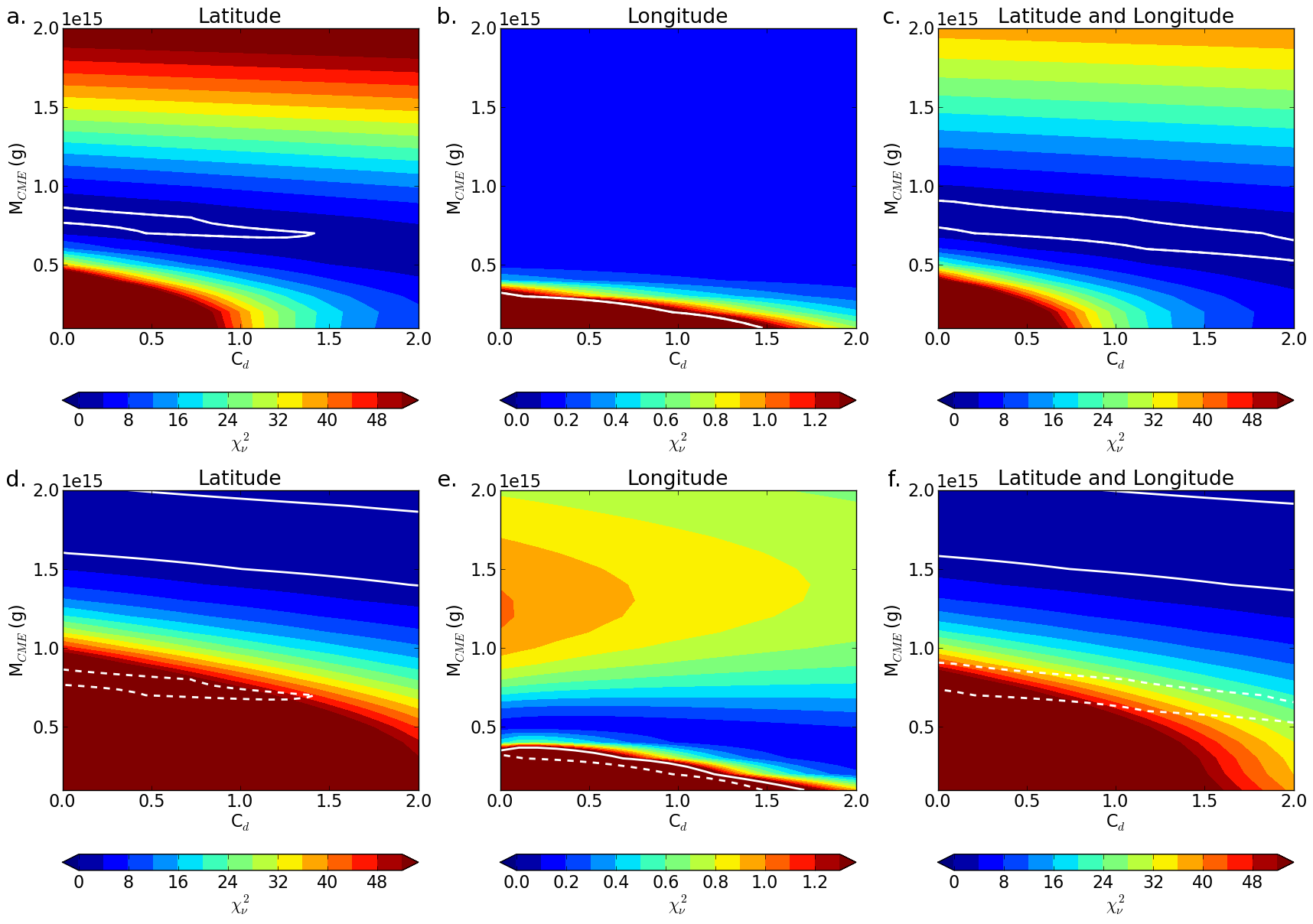}
\caption{Contours of the reduced chi-squared, $\chi^2_{\nu}$, versus mass and drag coefficient.  Panels (a)-(c) show results using the unscaled PFSS magnetic field model and Figs. \ref{fig:ps}(d)-(f) show results using the scaled PFSS magnetic field model.  Panels (a) and (d) show $\chi^2_{\nu}$ determined using only the latitude, Panels (b) and (e) show $\chi^2_{\nu}$ determined using only the longitude, and Panels (c) and (f) show $\chi^2_{\nu}$ determined using both latitude and longitude.  Note the difference in the contour range for the longitude panels.  The white line indicates where $\chi^2_{\nu}=1.5$ for all panels and in panels (d)-(f) the dashed white line corresponds to the region from the unscaled version.}\label{fig:ps}
\end{figure}

Figure \ref{fig:ps} shows contours of $\chi^2_{\nu}$ resulting from 400 ForeCAT simulations with masses between 10$^{14}$ g and 2x10$^{15}$ g and drag coefficients between 0 and 3.  In Fig. \ref{fig:ps}(a), the $\chi^2_{\nu}$ is computed using only the latitudinal points of G11 as they use the more commonly used technique to reconstruct the CME position.  In Fig. \ref{fig:ps}(b), the $\chi^2_{\nu}$ is computed using only the COR2 longitudinal points of G11.  Fig. \ref{fig:ps}(c) uses both the latitudinal and longitudinal points to get the total $\chi^2_{\nu}$ used to determine the best fit.  The white line corresponds to $\chi^2_{\nu}=1.5$.  A $\chi^2_{\nu}\approx 1$  indicates a good fit, but values significantly higher or lower than unity imply that the data is either under or overfitting the data.  

The latitudinal $\chi^2_{\nu}$ contours show an extended region which produces acceptable values of $\chi^2_{\nu}$ near unity.  These contours show some degeneracy between mass and drag coefficient with lower masses requiring higher drag coefficients.  

The longitudinal $\chi^2_{\nu}$ values in Fig. \ref{fig:ps}(b) includes points as low as 0.12 due to the low number of longitudinal points and their large uncertainty.  The longitude points alone cannot be used to constrain the CME input parameters.  

When the latitudinal and longitudinal points are combined into a single measure of $\chi^2_{\nu}$ the region corresponding to $\chi^2_{\nu}=1.5$ increases in Fig. \ref{fig:ps}(c) due to the low $\chi^2_{\nu}$ of the longitudinal comparison.  

We use only the latitudinal $\chi^2_{\nu}$ to determine the range of acceptable parameters.  By restricting the range of plausible input parameters to where $\chi^2_{\nu}\le 1.5$ we can restrict the CME mass to 7x10$^{14}$ to 8x$10^{14}$ g and a drag coefficient less than 1.4.  The best fit corresponds to 7.5x10$^{14}$ g for the mass and 0 for the drag coefficient.

Figure \ref{fig:ps}(d), (e), and (f), show the same as (a), (b), and (c), respectively, but for the scaled magnetic background.  All panels have the same contour levels of $\chi^2_{\nu}$ as the corresponding unscaled panels.  For low masses and small drag coefficients, the scaled background values of $\chi^2_{\nu}$ consistently exceed those of the unscaled background, however the majority of parameter space has a longitudinal $\chi^2_{\nu}$ below one.  The scaled background does produce $\chi^2_{\nu}\approx 1$ for the latitude for high masses, above 1.5x10$^{15}$ g.  This slightly exceeds the upper limit of $10^{15}$ g below 6 $\rsun$ determined by \citet{DeF13}.  However, it remains plausible that for the 2008 Dec 12 CME, which originated in the quiet sun, the magnetic field may decrease slightly less rapidly with distance than the PFSS model.

\subsection{Variation with Shape Parameters}\label{shapeparams}
\begin{figure}
\includegraphics[scale=0.4]{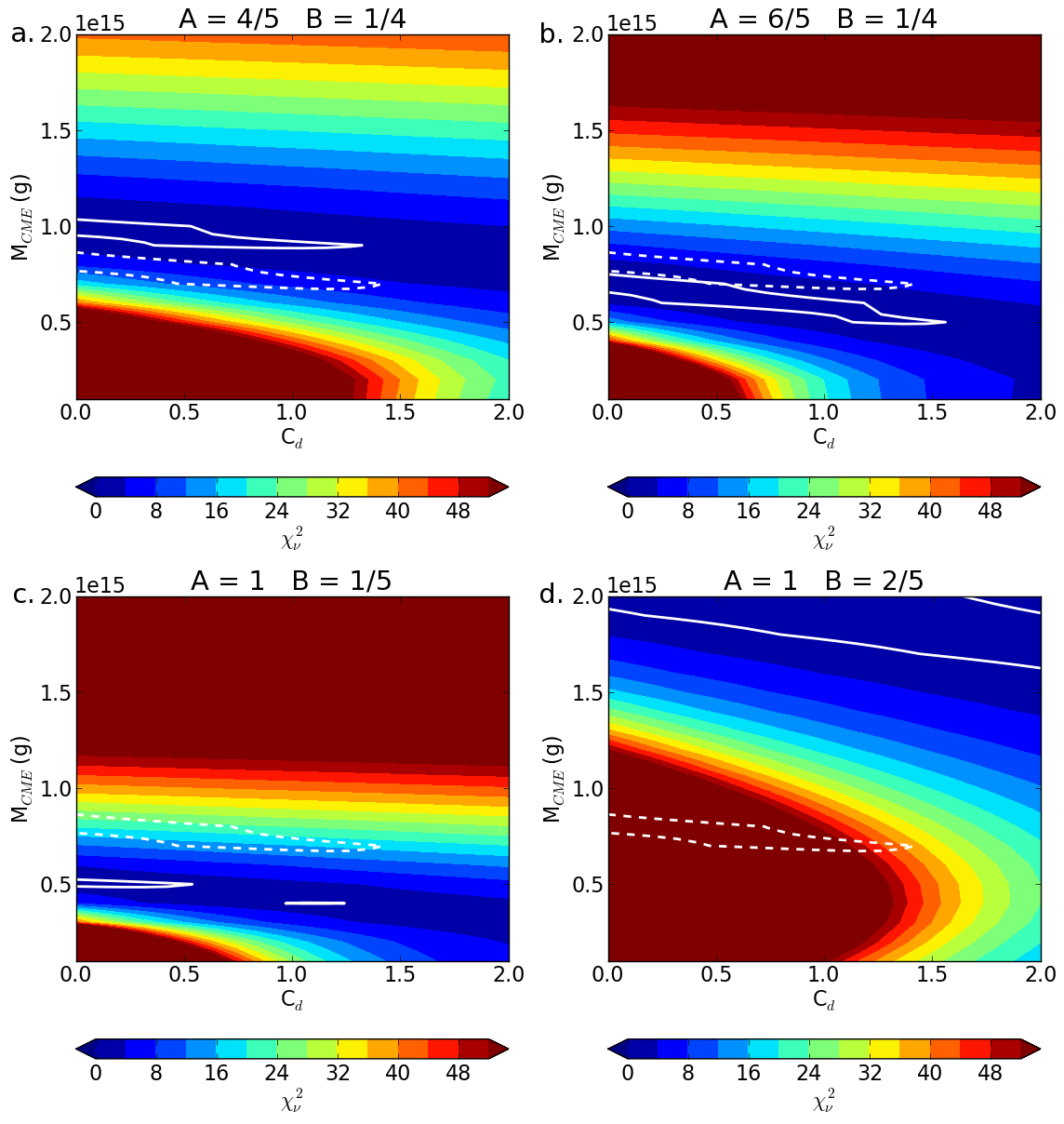}
\caption{Contours of the latitudinal reduced chi-squared, $\chi^2_{\nu}$, versus mass and drag coefficient (analogous to Fig. \ref{fig:ps}(a)).  Each panel uses a different set of shape ratios $A$ and $B$.  The solid white line indicates where $\chi^2_{\nu}=1.5$ for that $A$ and $B$ and the dashed white line shows $\chi^2_{\nu}=1.5$ from the case in Fig. \ref{fig:ps}(c).}\label{fig:AB}
\end{figure}

ForeCAT assumes that the CME height and cross-sectional width (parameters $a$ and $b$) maintain fixed ratios ($A$ and $B$) with the CME width.  The previous results used $A=1$ and $B=\frac{1}{4}$.  We vary $A$ and $B$ and compare the resulting $\chi^2_{\nu}$ contours with the first case.  Figure \ref{fig:AB} shows contours of the $\chi^2_{\nu}$ for only latitudinal points (analogous to Fig. \ref{fig:ps}(a)) for four different cases.  The solid white line shows the region corresponding to $\chi^2_{\nu}=1.5$ for that set $[A$,$B]$ and the dashed white line corresponds to $\chi^2_{\nu}$ from the control case of $[1$, $\frac{1}{4}]$ from Fig. \ref{fig:ps}.  These cases have longitudinal $\chi^2_{\nu}$ significantly below unity for most of mass and drag coefficient parameter space.

Fig. \ref{fig:AB} shows that changing the shape parameters either causes a slight difference in the range of $\chi^2_{\nu}$ near unity or results in a poor fit to the data such that $\chi^2_{\nu}$ never reaches unity.  When the CME height is decreased relative to the width (Fig. \ref{fig:AB}(a)) the range of acceptable masses and drag coefficient decreases, and shows a shift toward higher masses.  The decrease in height causes these CMEs to experience stronger magnetic forces so the range of acceptable parameters tends toward higher masses, as high as 10$^{15}$ g.  The range of acceptable drag coefficient decreases.  When the CME height is increased (Fig. \ref{fig:AB}(b)), the range in parameter space yielding acceptable $\chi^2_{\nu}$ shifts toward smaller masses, with acceptable values as low as 5x10$^{14}$ g.  The increase in height causes these CMEs to initially experience weaker magnetic forces so only low mass cases can reproduce the extensive observed deflection.  

Decreasing the cross-sectional radius (Fig. \ref{fig:AB}(c)) causes a decrease in the range of acceptable parameters which is shifted toward lower mass, about 4.5x$10^{14}$ g.  If the cross-sectional width decreases further the range corresponding to $\chi^2_{\nu}=1.5$ continues to decrease and eventually disappears.  Increasing the cross-sectional radius results in a significant shift of the region corresponding to $\chi^2_{\nu}=1.5$ to masses larger than 1.7x10$^{15}$ g, suggesting that the observations cannot be reproduced with a cross-sectional width much larger than that used in Fig. \ref{fig:ps}(c).  We find that despite not being able to accurately determine several CME shape parameters we can still constrain the CME mass and solar wind drag coefficient.

\section{Conclusion}
The ForeCAT results reproduce the observations of the 2008 December 12 CME and by computing the reduced chi-squared, $\chi^2_{\nu}$, we can constrain the CME mass and drag coefficient.  The CME shape is only constrained, rather than uniquely determined from observations.  CME masses between 4.5x10$^{14}$ and 1x$10^{15}$ g and drag coefficients between 0 and 1.4 correspond to good fits with $\chi^2_{\nu}$ near unity.  We find that the observations can also be reproduced with a magnetic background that decreases slightly less rapidly with distance than the PFSS model.  This comparison shows promise that ForeCAT can be successfully used to predict CME deflections or constrain CME properties (mass), or probe the solar background (drag coefficient and magnetic field).  Future development will be done to improve the expansion and propagation models so that they do not need to be predetermined from observations.    

\acknowledgements
The authors thank J. P. Byrne for providing the CME observations.  This work was supported by grants from the Coordination for the Improvement of Higher Education Personnel (Capes - Brazil, 99999.016405/2012-09; to L.F.G.d.S.)  The authors would like to thank the anonymous referee for his comments.

\end{document}